\documentclass[aps,prl,preprint,superscriptaddress, nobibnotes]{revtex4}

\usepackage{graphicx}
\usepackage{dcolumn}
\usepackage{bm}
\usepackage{amssymb}
\usepackage{color}
\usepackage{amsmath}
\usepackage{epstopdf}
\usepackage{natbib}
\usepackage{float}

\graphicspath{{figures/}}
\begin{document}

	%\preprint{AIP/123-QED}

\title[Feedback-controlled microbubble generator producing one million monodisperse bubbles per second]{Feedback-controlled microbubble generator producing \\ one million monodisperse bubbles per second}
% Force line breaks with \\

\author{Benjamin van Elburg}
\thanks{These authors contributed equally to this work.}
\affiliation{ 
	Physics of Fluids Group, Technical Medical (TechMed) Center and MESA+ Institute	for Nanotechnology, University of Twente, P.O. Box 217, 7500 AE Enschede, The Netherlands.%\\This line break forced with \textbackslash\textbackslash
}%

\author{Gonzalo Collado Lara}% 
\thanks{These authors contributed equally to this work.}
\affiliation{ 
	Physics of Fluids Group, Technical Medical (TechMed) Center and MESA+ Institute	for Nanotechnology, University of Twente, P.O. Box 217, 7500 AE Enschede, The Netherlands.%\\This line break forced with \textbackslash\textbackslash
}%
\affiliation{ 
	 Biomedical Engineering, Thoraxcenter, Erasmus MC, Rotterdam, The Netherlands.%\\This line break forced with \textbackslash\textbackslash
}%

\author{Gert-Wim Bruggert}
\affiliation{ 
	Physics of Fluids Group, Technical Medical (TechMed) Center and MESA+ Institute	for Nanotechnology, University of Twente, P.O. Box 217, 7500 AE Enschede, The Netherlands.%\\This line break forced with \textbackslash\textbackslash
}%

\author{Tim Segers}
\affiliation{ 
	Physics of Fluids Group, Technical Medical (TechMed) Center and MESA+ Institute	for Nanotechnology, University of Twente, P.O. Box 217, 7500 AE Enschede, The Netherlands.%\\This line break forced with \textbackslash\textbackslash
}%
\affiliation{BIOS Lab-on-a-Chip group, Max-Planck Center Twente for Complex Fluid Dynamics, MESA+ Institute for Nanotechnology, University of Twente, Enschede, Netherlands}
\author{Michel Versluis}
\affiliation{ 
	Physics of Fluids Group, Technical Medical (TechMed) Center and MESA+ Institute	for Nanotechnology, University of Twente, P.O. Box 217, 7500 AE Enschede, The Netherlands.%\\This line break forced with \textbackslash\textbackslash
}%
\author{Guillaume Lajoinie}
\affiliation{ 
	Physics of Fluids Group, Technical Medical (TechMed) Center and MESA+ Institute	for Nanotechnology, University of Twente, P.O. Box 217, 7500 AE Enschede, The Netherlands.%\\This line break forced with \textbackslash\textbackslash
}%

\date{\today}% It is always \today, today,

\begin{abstract}
Monodisperse lipid-coated microbubbles are a promising route to unlock the full potential of ultrasound contrast agents for medical diagnosis and therapy. Here, we present a stand-alone lab-on-a-chip instrument that allows microbubbles to be formed with high monodispersity at high production rates. Key to maintaining a long-term stable, controlled, and safe operation of the microfluidic device with full control over the output size distribution is an optical transmission-based measurement technique that provides real-time information on production rate and bubble size. We feed the data into a feedback loop and demonstrate that this system can control the on-chip bubble radius (2.5 to 20 $\mu$m) and the production rate up to $10^6$ bubbles/s. The freshly formed phospholipid-coated bubbles stabilize after their formation to a size approximately two times smaller than their initial on-chip bubble size without loss of monodispersity. The feedback control technique allows for full control over the size distribution of the agent and can aid the development of microfluidic platforms operated by non-specialist end users.
\end{abstract}

\maketitle

\section{\label{sec:level1}Introduction}

Contrast-enhanced ultrasound imaging relies on the use of an ultrasound contrast agent (UCA), comprising a suspension of stabilized microbubbles, that is injected intravenously into the circulation to enhance the echogenicity of the blood pool.~\cite{Lindner2004} Microbubbles undergo volumetric oscillations when exposed to ultrasound, thereby generating strong harmonic echoes.~\cite{Averkiou2020} Key to their efficient use as a contrast agent is the compressibility of the microbubble gas core, which allows microbubbles to resonate at the typical ultrasound frequencies used for medical imaging (1-10 MHz).~\cite{Leighton1994} When a bubble is driven at its resonance frequency, the relative volumetric oscillation amplitude is at maximum, resulting both in a maximized nonlinear echo response and a maximized mechanical effect on neighboring tissues.~\cite{Kooiman:2014zr,Versluis2020} The characteristic eigenfrequency of a bubble is, however, intimately linked to its size,~\cite{Minnaert1933} while commercial UCAs consist of a polydisperse suspension of lipid-coated microbubbles with diameters ranging from 1 to 10~$\mu$m.~\cite{Frinking2020} As a result, only a small fraction of the contrast microbubbles resonates to the driving ultrasound field,~\cite{segers2016uniform} limiting the sensitivity of contrast-enhanced ultrasound imaging and molecular imaging with ultrasound,~\cite{Klibanov2006, AbouElkacem2015, Streeter2010} as well as the efficacy of theranostic applications. Control over the size distribution of UCAs~\cite{segers2018monodisperse,Helbert2020}  may therefore open a route to unlock the full potential of novel medical applications of microbubbles, including drug and gene delivery,~\cite{Tsutsui2004, Hernot2008, Carson2012, Dewitte2015} blood-brain-barrier opening,~\cite{Hynynen2006,Burgess2015} sonothrombolysis,~\cite{Molina2009} and sonoporation,~\cite{Bourn2020,Helfield2016} that all rely on the volumetric oscillation amplitude of the bubbles.~\cite{Roovers2019,Stride2020}

Microbubble suspensions with a narrow size distribution can be obtained from a native polydisperse size distribution by mechanical filtration,~\cite{Emmer2009} decantation,~\cite{Goertz2007d} centrifugation,~\cite{Feshitan2009} and microfluidic acoustic or size-sorting.~\cite{Segers2014,Kok2014} Microfluidics can furthermore be applied to produce highly concentrated (10$^{10}$~bubbles/mL) bubble suspensions~\cite{Peyman2012} and sub-micron bubbles.~\cite{Peyman2016} Monodisperse bubbles can also be directly synthesized  in a microfluidic flow-focusing device.~\cite{Ganan-calvo2001a} Here, a gas thread is focused between an aqueous co-flow through a constriction where the gas phase destabilizes and pinches off to release monodisperse bubbles.~\cite{Anna2003,Garstecki2004} The lipid coating material is typically dispersed in the aqueous phase~\cite{Hettiarachchi2007, Talu2008, Shih2013,Parrales2014,Gong2014} from which it can adsorb onto the gas-liquid interface.~\cite{Segers2016,Segers2017} Recent advances in the field of lipid-coated microbubble formation by flow-focusing allow the synthesis of foam-free,~\cite{Segers2020} highly concentrated ($>$ 200~million bubbles/mL),~\cite{Segers2019} and monodisperse bubble suspensions with a high stability both in the collection vial and upon dilution.~\citep{Segers2017,Segers2019} These developments pave the road to translation from a scientific method to a (high-volume) industrial production or (clinical-grade) bedside production facility for medical bubbles. 

Clinical and industrial translation both require long-term precise control over bubble size distribution and production rate together with a continuous monitoring of the on-chip bubble formation process to guarantee the quality and clinical safety of the product. However, flow-focusing devices are typically prone to clogging and a slow drift in bubble size, requiring microfluidic expertise, whereas clinical and industrial translation require a user-friendly system accessible to non-experts. Microfluidic bubble synthesis platforms will thus greatly benefit from a simple, safe, and reliable feedback system that allows adaptive flow control to maintain a constant bubble size and production rate. This system can then stabilize the bubble production against changes in flow conditions due to, e.g., a passing dirt particle that escaped upstream on-chip filtering, resulting in increased flow resistance or partial clogging. Furthermore, changing the wetting conditions of the channel walls results in a noticeable change in the contact line position and curvature of the gas jet in the gas inlet channel. Finally, using a feedback system will allow the non-expert user with limited knowledge about the operating range of the employed chip to simply set a bubble size and a production rate. 

Feedback control of microfluidic droplet and bubble generators has frequently been reported in literature.~\cite{Miller2010,Zeng2015,Dekker2018,Rodriguez2008,Rickel2018,Gong2008,Shih2011} Digital feedback systems rely on the ability to accurately detect the formed bubbles and droplets. Reported detection methods for lab-on-a-chip devices include Coulter Counter-like electrical sensing with integrated electrodes~\cite{Dekker2018,Rodriguez2008,Rickel2018,Gong2008,Shih2011,Nguyen:2013qa,Mansor:2015mi} and optical sensing, either using integrated on-chip optical waveguides~\cite{Nguyen2006} or using an external light source.~\cite{Rabaud2011} Furthermore, (high-speed) imaging together with digital image processing has been employed as a detection system for feedback-controlled microfluidic droplet generators.~\cite{Miller2010,Zeng2015,Vo2017} However, the reported feedback systems were operated at kHz production rates whereas the present application requires MHz detection and processing rates. Recently, \citet{Xie2020} demonstrated feedback control of a flow-focusing device using electrical on-chip sensing to dynamically monitor and regulate the produced microbubble radius between 7~and 12~$\mu$m at a maximum rate of 10$^5$~bubbles/s with a proportional-integral feedback control system.~\cite{Xie2020} Electrical sensing methods are fast but present the inconvenience of requiring a chip design that includes electrodes, thereby increasing the cost and technical challenge involved in the production of the chip. Moreover, electrical sensing methods require a precise calibration in terms of electrical signal versus bubble size that has to be performed for each type of liquid and channel geometry. In contrast, a bubble detector placed outside of the lab-on-a-chip device would increase the versatility of the detection system and it lowers the production cost of the chip since it can be operated using various chips and device geometries.

Here, we propose a simple and fast optical detection method and feedback system based on the decreased transmission of light through the chip as newly formed microbubbles cross a low-intensity focused laser beam. We show that this optical system can measure both the production rate and the size of microbubbles produced in a microfluidic flow-focusing device in real-time. Production rates exceeding one million bubbles per second were readily measured for clinically relevant bubble radii below 5~$\mu$m. We also show that this system can be used as an input to a feedback system to control and stably operate a bubble generator chip. Finally, we show that this system is valuable for the expert user by offering a full characterization of the performance of the lab-on-a-chip device much faster than any other system known to the authors. This includes the possibility to monitor transient states of the device.

\begin{figure}[]
	\centering
	\includegraphics[width = 0.6\columnwidth]{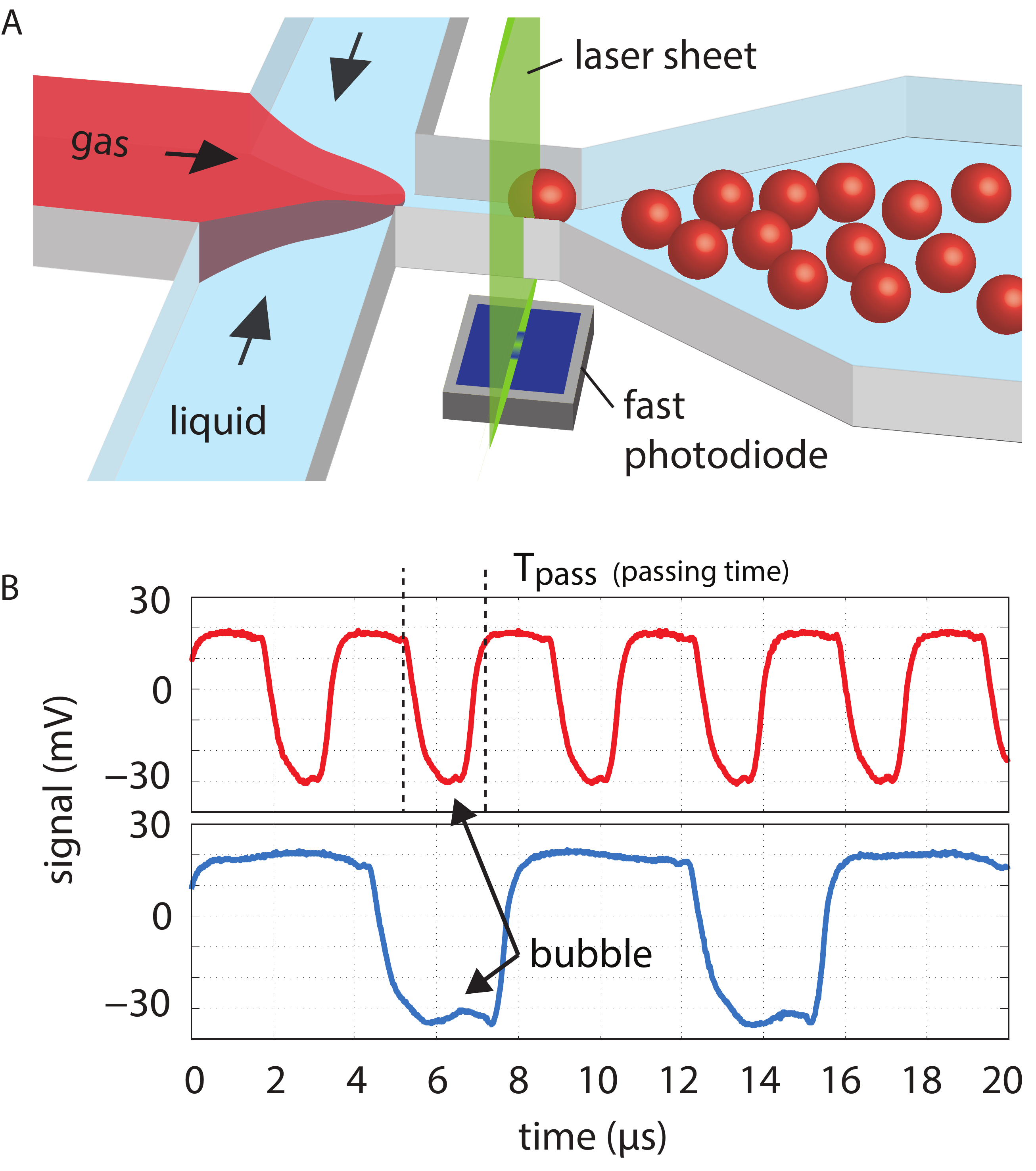}
	\caption{(A) Schematic representation of the optical bubble detection setup. A laser sheet is formed using a cylindrical lens and positioned at the end of the bubble formation channel. A fast reversed-biased photodiode was employed to measure the change in light intensity observed when bubbles pass through the laser sheet. (B) Typical output voltage of the photodetector for different bubble production rates resulting in different passing times.}
	\label{fig:setup}
\end{figure}

\section{Methods}
The performance of the system was measured using a generic flow-focusing device with a narrow bubble formation channel as introduced by~\citet{Castro-Hernandez:2011dz} Here, the chip was manufactured in glass with a bubble formation channel length $L=30~\mu$m, as described by~\citet{Segers2018} This flow focusing chip geometry can produce on-chip bubble sizes ranging from 2.5~to 20~$\mu$m in radius.

The bubble detection system consists of a low power (5~mW) laser module of which the elliptical beam was focused into a laser sheet by a set of two cylindrical lenses, a spherical lens, and a 50$\times$ magnification microscope objective. The combination of lenses allows to control the width and thickness of the laser sheet in order to maximize the SNR of the detection system. The sheet geometry (Fig.~\ref{fig:setup}A) provides a robust alignment with the bubble stream. The transmitted light was received by a fast photodiode (FDS100, Thorlabs, Si Detector, 350~-~1100~nm, 14~ns rise time, 13~mm$^2$ aperture) connected to a 100~$\Omega$ load to optimize the trade-off between sensitivity and response time. The output voltage of the diode was measured with a USB oscilloscope connected to a computer running Matlab to download the recorded waveforms ($\sim$15 waveforms/s). The production rate was calculated from the stored waveforms using a cross-correlation method (3--5 times faster than a Fourier transform method).
The passing time (Fig.~\ref{fig:setup}B) of the bubbles was measured from the waveforms using a slope-corrected thresholding and peak detection method (see Appendix~B).

In the first set of experiments, the liquid consisted of demineralized water mixed with a surfactant (Tween 80, Sigma Aldrich). The flow was controlled by a mass-flow controller (EL-Flow series, Bronkhorst, NL) with a typical flow rate of 10~mL/h. The gas flow (typically 15~mL/h) was also mass-flow controlled and the gas pressure was measured in the gas supply tube by a pressure sensor (Digitron).
The mass-flow controllers were operated by a computer running tailor-made software from which it was possible to program fully automated sequences of mass-flow rates. The actual flow rates were measured 10~times per second from the mass-flow controller I/O ports. 

\begin{figure}[t]	
	\centering
	\includegraphics[width = 0.6\columnwidth]{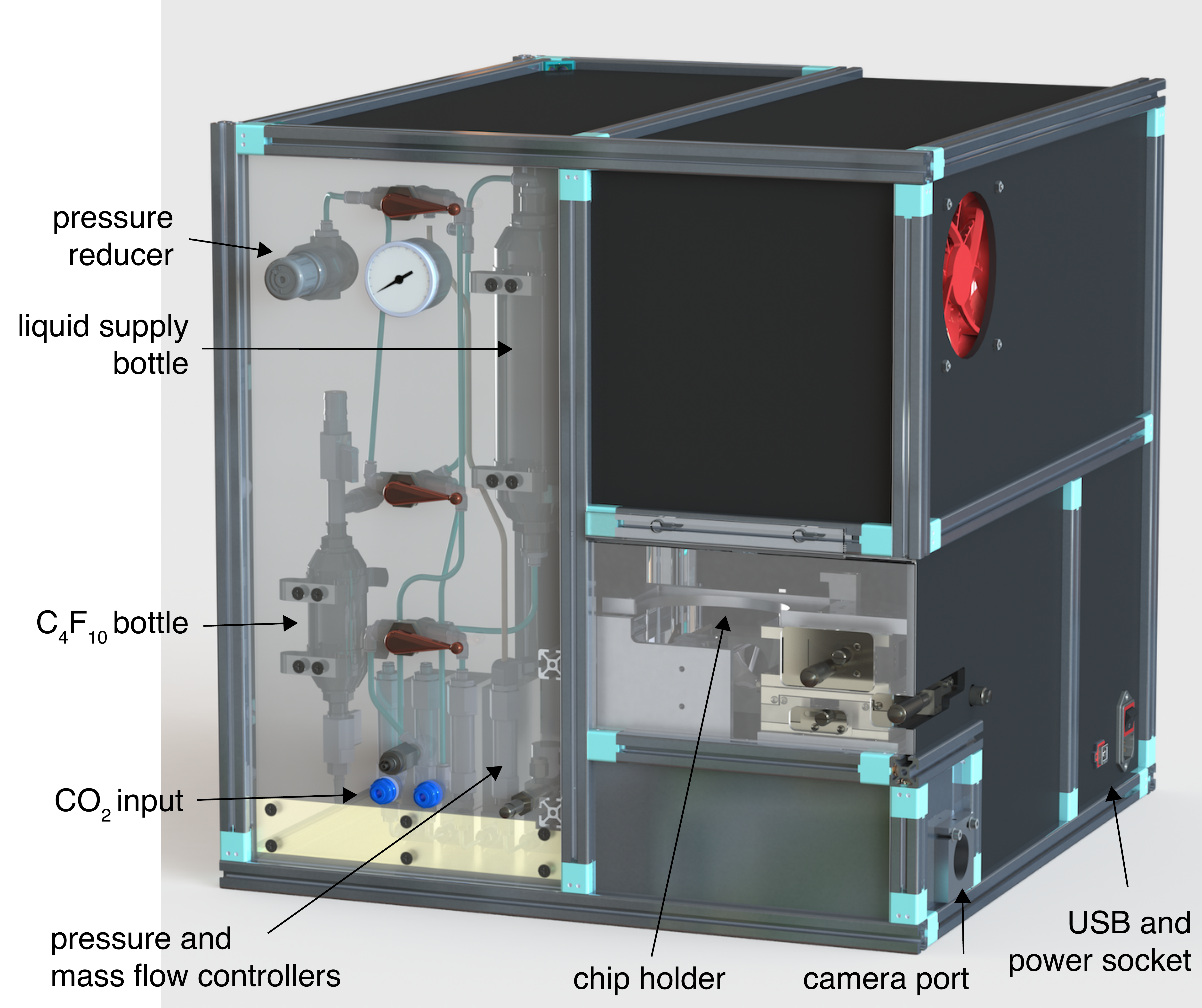}
	\caption{CAD drawing of the stand-alone monodisperse bubble production system.}
	\label{F:1a}
\end{figure}
For the final set of experiments, a pressure controller (EL-PRESS series, Bronkhorst, NL) was added to the initial setup to decrease the response time of the system. The pressure controller was included in the gas line in place of the mass-flow controller for the gas. The final setup was built as a stand-alone system in a self-contained box, embedding the mass-flow controllers, the gas pressure controller connected to and the optical components, which included a laser, laser optics, microscope objective, imaging lenses, and a USB-connected CCD camera (Lumenera LM165M). 
A CAD drawing of the stand-alone system is shown in Fig.~\ref{F:1a}.  
In this experiment, we demonstrate the formation of monodisperse phospholipid-coated microbubbles. Here, we use an aqueous phospholipid mixture comprising DSPC and DPPE-PEG5000 mixed at a 9:1 molar ratio, as in ~\cite{Segers2019}, at a total concentration of 12.5 mg/mL and prepared as described by ~\citet{Segers2017} Microfluidically-formed lipid-coated bubbles decrease in size after formation by a factor of 2 to 3 until their stable size is reached.~\cite{Segers2017, Segers2019} To allow for the efflux of the microbubble filling gas without large foam bubble formation through Ostwald ripening, the bubbles were filled with a multi-gas-component mixture of a gas with a high aqueous solubility and a gas with a low aqueous solubility. Here we employed a gas mixture of 12 vol\% of C$_4$F$_{10}$ in CO$_2$. As demonstrated before, using the present gas mixture, the gas in the stable bubbles is composed of nearly pure C$_4$F$_{10}$ due to the rapid efflux of the CO$_2$ gas during bubble stabilization.~\cite{Segers2020} The mixture was fed to the pressure controller at a stable pressure by using two mass flow controllers and a pressure sensor. The C$_4$F$_{10}$ mass flow was dynamically adjusted by a closed-loop controller to ensure a stable mixing ratio.

Waveforms were acquired using a digital storage oscilloscope (Picoscope, 3403D). Ambient temperature in the stand-alone system was controlled and stabilized at 55$^\circ$C to allow for the coalescence-free production of phospholipid-coated bubbles containing gaseous perfluorobutane (C$_4$F$_{10}$) for increased stability against dissolution.~\cite{Borden2018}

The elevated temperature dramatically reduces the on-chip bubble coalescence probability.~\cite{Segers2019} Another advantage of the higher operating temperature is the corresponding increased driving pressure of the C$_4$F$_{10}$ gas to 5 bars.~\cite{Segers2020} The increased driving pressure allows monodisperse bubble production at high production rates, not only in the present flow-focusing device where we used a maximum gas pressure on the order of 2 bars, but also in devices with a higher pressure drop, e.g., due to an increased length or reduced cross sectional area of the narrow channel in which the bubbles are formed. Controlled heating was achieved using an Arduino Nano, programmed to control a heating element via a 4-20~mA power regulator (SO445420, Celduc). The temperature in the self-contained system was kept homogeneous using a set of fans. The stand-alone system further embeds a power supply and is linked to the control computer via a single USB cable.

\section{Results and discussion}
\subsection{Transients}
\begin{figure}[b]
	\centering
	\includegraphics[width = 0.75\columnwidth]{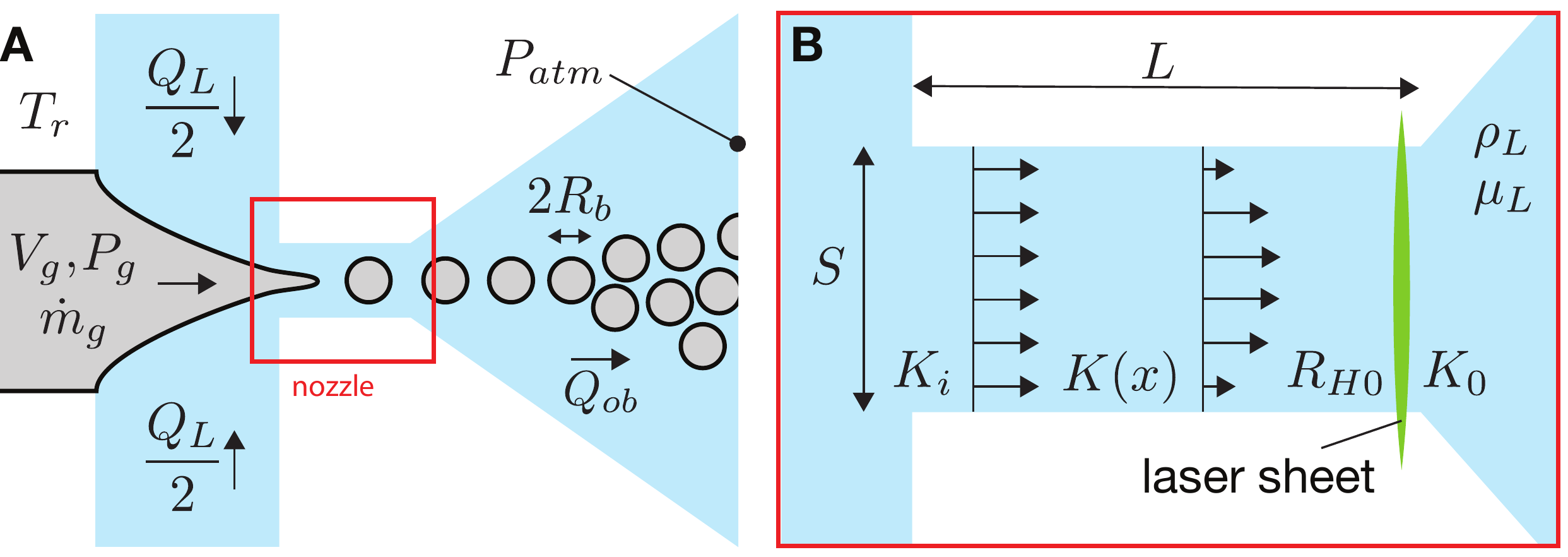}
	\caption{(A) Schematic overview of the microfluidic device with the relevant physical parameters. (B) Magnification of the channel area in which the bubbles are formed with the loss parameters $K$ and $R_H$ accounting for the pressure drop. The laser sheet, used for counting and sizing the bubbles, is positioned at the end of the narrow bubble formation channel.}
	\label{fig:Parameters}
\end{figure}
In the following, we will consider two sets of experiments producing microbubbles. In the first, the liquid flow rate was changed in steps while maintaining a constant gas flow rate. In the second, the liquid flow rate was kept constant and the gas flow rate was varied. For each change in flow rate, we recorded the pressure change in the gas line as a function of time.

Using simple fluid dynamics considerations detailed in Appendix A, we can show that a gas-flow-rate-controlled chip driven at high pressure presents a typical transient time: 
\begin{equation} \label{eq:tauP}
\tau = \frac{1}{2} \frac{V_g \left(R_{H0} + K_2 Q_L\right)}{P_{atm}}
\end{equation}
for the pressure build-up.
Note that the prefactor $1/2$ becomes 1 in case of low flow rates (low pressure driving limit). The response to a pressure change then writes:
\begin{equation} \label{eq:app_pressure}
P_{g}\left(t\right) \approx P_{g,t=0} + \left(P_{g,eq} -P_{g,t=0} \right)\left( 1-e^{-t/\tau} \right),
\end{equation}
with
\begin{equation}\label{eq:app_pressure_eq}
P_{g,eq}  = \frac{1}{2} \left( \kappa + \sqrt{\kappa^2 + 2 \left(R_{H0} + K_2  Q_L\right) Q_L P_{atm}}\right),
\end{equation}
and where
\begin{equation}\label{eq:delt}
\kappa = P_{atm}+\left(R_{H0} + K_2 Q_L\right)\left( Q_L + \frac{T_r R_g \dot{m}_g}{P_{atm} \tilde{M_g}}\right).
\end{equation}
In addition, these considerations show that the gas flow rate experiences the same transient as the gas pressure. Eq.\ref{eq:app_pressure_eq} in Appendix A shows that the gas flow rate depends in a non-linear way on the variations of the liquid flow rate. Thus, a constant input gas flow rate does not necessarily imply a constant outlet gas flow rate.

\begin{figure}[t]
	\centering
	\includegraphics[width = 0.8\columnwidth]{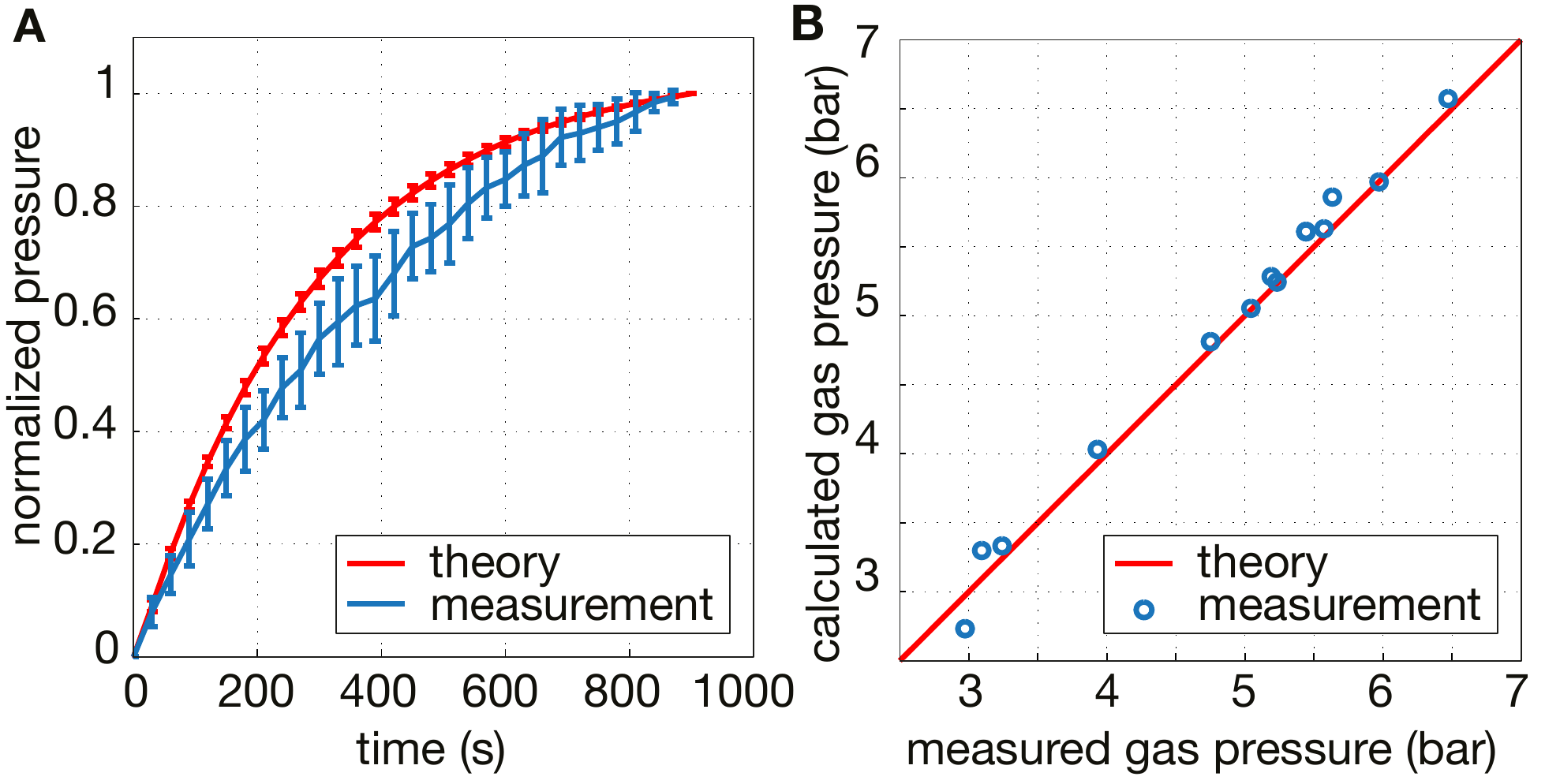}
	\caption{Pressure in the gas line. (A) Variation of the pressure in the chip during the transient time. Experimental reading (blue curve) and theory (red curve) are in good agreement. The measured pressure rise is slightly slower than the simulated one. (B) Comparison between the calculated and measured pressures.}
	\label{fig:pressure}
\end{figure}
Figure~\ref{fig:pressure}A compares the measured gas pressure to the calculated gas pressure from the model. In Fig.~\ref{fig:pressure}A, the pressure steps are normalized to 1 and plotted as a mean. The error bars represent the standard deviation. The standard deviation in the theoretical curve represents the spread in responses to different pressure steps. The transients estimated from the simulation and the measured ones are in good agreement. The slower pressure change observed in experiment can be attributed to an underestimation of the dead gas volume (0.8 mL in our setup). A full comparison of the simulated pressure with the measured one is displayed in Fig.~\ref{fig:pressure}B and here again, good agreement is observed.

\subsection{Bubble counting}

To perform the bubble counting, the laser beam was positioned at the end of the narrow bubble formation channel (see Fig.~\ref{fig:Parameters}B). Each bubble passing through the beam results in a drop in transmitted laser light. Example waveforms are shown in Fig.~\ref{fig:setup}B.

\begin{figure}[htb]
	\centering
	\includegraphics[width = 0.71\columnwidth]{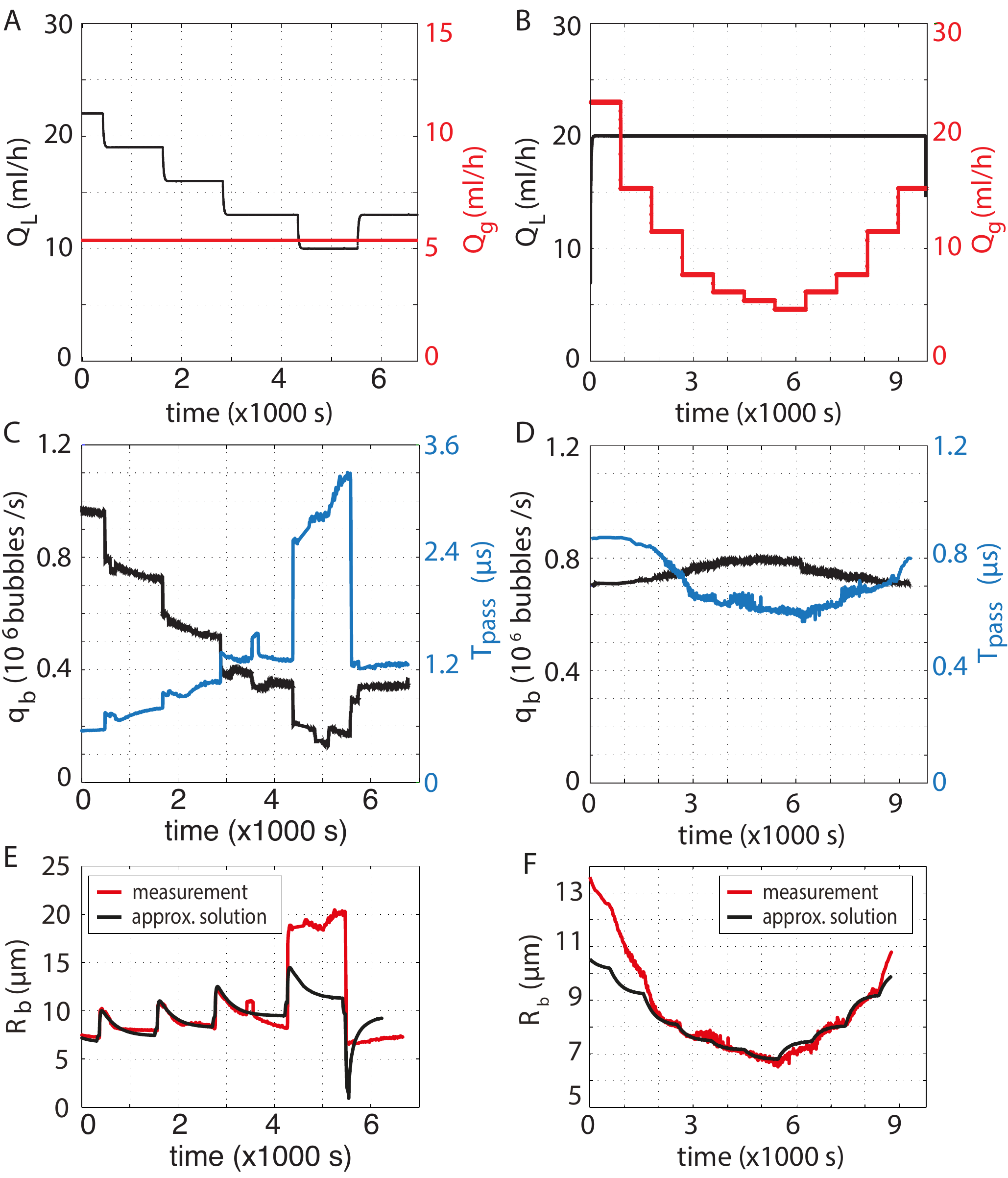}
	\caption{Measured inputs and outputs of the chip. The inputs in (A) and (B) are the liquid flow rate $Q_L$ and the gas flow rate $Q_g$, respectively. In (A) the liquid flow rate was varied while the gas flow rate was kept constant. In (B) the gas flow rate was varied while the liquid flow rate was kept constant. The directly measured outputs in (C) and (D) are the production rate $q_b$ and the passing time $T_{pass}$ of the formed bubbles. (E) and (F) show the measured bubble size (red) and the analytically calculated bubble size (black) as a function of time for the varying liquid flow rate, as in (C), or gas flow rate as in (D). (G) shows the microfluidic device during microbubble production. (H) A diluted monodisperse bubble sample.}
	\label{fig:IO}
\end{figure}

The results of two experiments where either the liquid flow rate (Figs.~\ref{fig:IO}A, C, E) or the gas flow rate (Figs.~\ref{fig:IO}B, D, F) was varied are plotted in Fig.~\ref{fig:IO}. Figure~\ref{fig:IO}C and D display the measured production rate $q_b$ and passing times $T_{pass}$ of the bubbles. From the production rate and the corresponding passing time we can estimate the bubble size (see Sec. \ref{Bubble_sizing}). The obtained bubble radius over time is plotted in Figs.~\ref{fig:IO}E and~F. In order to prove the fast response time and the reliability of the system, the chip was driven in extreme conditions in terms of gas pressure and liquid flow rate. This resulted in a bubble production rate up to $10^6$ bubbles/s. 

Higher production rates are possible but were observed to have a negative effect on the production stability of the chip. At production rates exceeding $1.2\times10^6$ bubbles/s, an unstable position of the contact line of the gas jet was observed meaning that the gas jet pinning line jumped back and forth resulting in unstable bubble production. This instability may also be the result of the formation of cavitation bubbles and resulting pressure fluctuations in the expanding outlet of the chip at these high production rates.

\begin{figure}[t]
	\centering
	\includegraphics[width = 0.5\columnwidth]{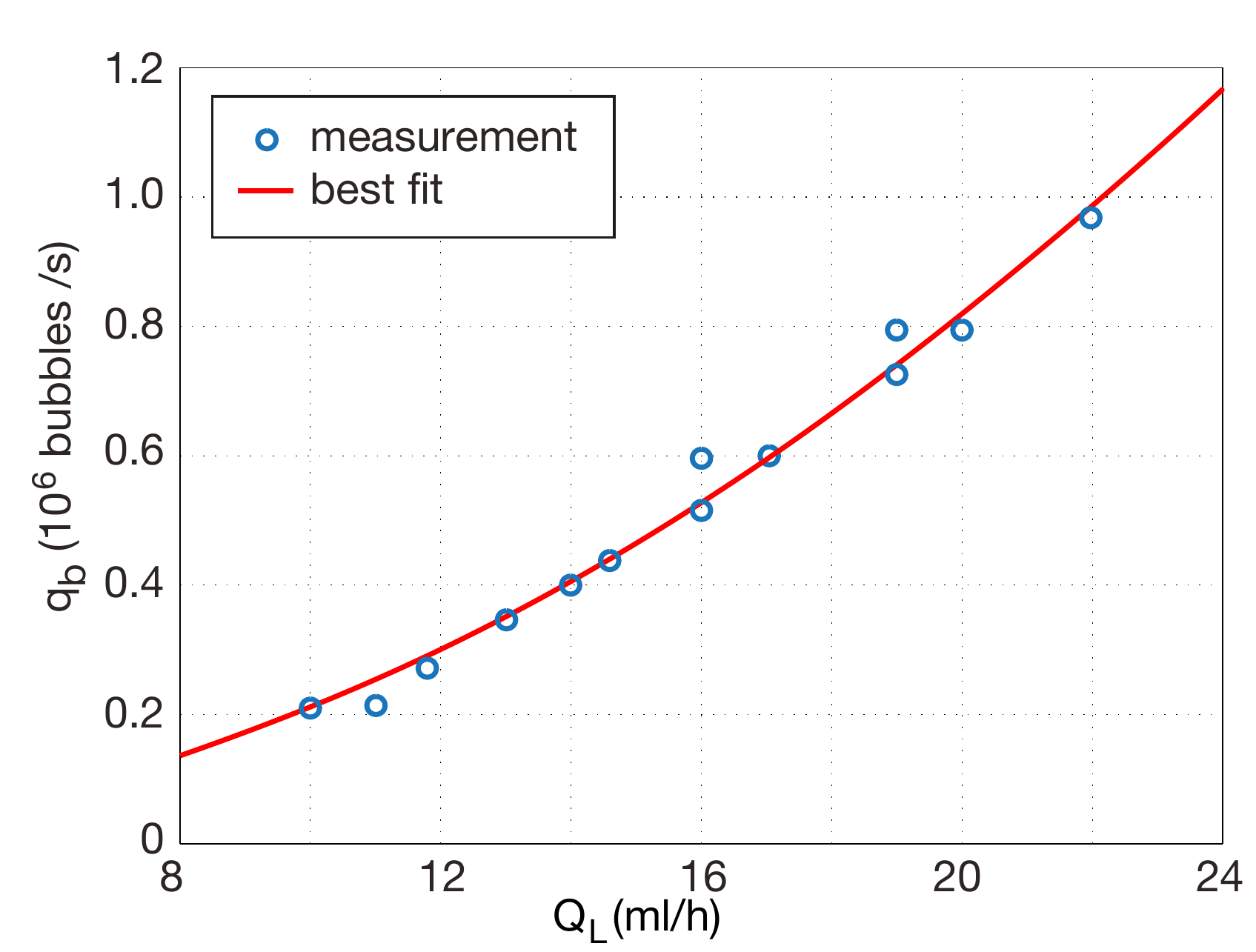}
	\caption{Measured production rate $q_b$ of the microbubbles versus the liquid flow rate $Q_L$. The experimental data is displayed as the blue dots and the best fit is depicted by the solid red line. The best fit is given by: $q_b = 2.313 Q_L^{1.96}$, with a coefficient of determination of $R^2$~=~0.99.}
	\label{fig:freq_and_bub_si}
\end{figure}

From Fig.~\ref{fig:IO}C and D, it can be observed that the liquid flow rate has a strong influence on the bubble production rate whereas the gas pressure does not. The production rate in Fig.~\ref{fig:IO}C quickly follows the change in liquid flow rate, followed by smaller variations occurring on the timescale of the pressure transient. In Fig.~\ref{fig:IO}D the production rate only varies by a few percent whereas the gas flow rate varies by a factor 5. These detailed relations provide valuable insight in the fluid dynamics governing the bubble production. For consistency with the concern of generality and simplicity, we perform a straightforward scaling with the liquid flow rate, as shown in Fig.~\ref{fig:freq_and_bub_si}. This scaling gives a good first-order result and can be implemented in a computer model to give a direct prediction for the production rate from the liquid flow rate. Note that the measurement can be readily repeated on any different chip geometry. 

\subsection{Bubble sizing \label{Bubble_sizing}}
The bubble radius $R_b$ is the second parameter required to control and characterize the bubble production in real time. In contrast to the production rate, this quantity is not directly measured. The bubble radius is a function of both the passing time of the bubble $T_{pass}$ (Fig.~\ref{fig:setup}B) measured from the waveforms, and the bubble velocity. 
In the chip presently investigated, the flow is not fully developed at the end of the bubble formation channel (see Appendix B) where the laser sheet was positioned. Therefore:
\begin{equation} \label{eq:bub_si}
R_b = - \dfrac{D_{laser}}{2} + \beta T_{pass} \frac{Q_{g}+Q_L}{S},
\end{equation}
where $\beta$ is a flow profile parameter that accounts for the undeveloped laminar flow. $\beta$ can vary between $\beta = 1$ for a parabolic profile and $\beta = 1/2$ for a flat profile
$D_{laser}$ is the thickness of the laser sheet, $T_{pass}$ is the passing time extracted from the waveforms, and $Q_L$ from the liquid mass flow rate measured directly by the mass-flow controller. $Q_g$ follows from the set gas mass flow rate taking into account the gas pressure (see Appendix A).
Finally, $D_{laser}$ can be estimated from the diffraction limit:\cite{Born1999}
\begin{equation} \label{eq:las_diam}
\dfrac{D_{laser}}{2} \approx 1.22\frac{\lambda F}{D} \approx 0.8~\mu \text{m},
\end{equation}
Here, $\lambda = 405$ nm is the wavelength of the laser, $F$ is the working distance of the objective and $D$ the diameter of the beam entering the microscope. Note that the minimum detectable bubble size is thus given by the laser sheet thickness, which in turn is governed by the diffraction limit. This result can easily be translated to droplet sizing by considering the inner flow to be incompressible, which reduces the prefactor of Eq.~\ref{eq:phioutandB} to 1 since in that case $\bar{Q}_{g} = Q_{g}$.

The measured and calculated bubble sizes are shown in Fig.~\ref{fig:IO}E and F, for the constant gas flow rate case (Fig.~\ref{fig:IO}A) and for the constant liquid flow rate case (Fig.~\ref{fig:IO}B), respectively. We find good agreement between the experimental and calculated curves. Two main discrepancies can be pointed out that correspond to a breakdown of the validity of our generic model. The first one is visible in Fig.~\ref{fig:IO}E when the bubble radius reaches a size of 20~$\mu$m. In this case, the gas flow rate to liquid flow rate ratio is large and the influence of bubble compressibility on the flow can no longer be neglected (as per our assumption). On the other hand, this flow rate ratio regime, producing larger bubbles, is of lesser interest for our clinical application. The second discrepancy appears in Fig.~\ref{fig:IO}F at early times. The gas flow rate is then dropping rapidly from a very large initial value, which is a result from a startup transient when the system is switched on before t=0. While the transient pressure  recovery lasts for the first 1500~s, note that the measured bubble size remains accurate. In practice, this difficulty disappears if a feedback loop is implemented, as demonstrated in Sec.~ \ref{sec:opti}. 

\section{Performance optimization and feedback loop \label{sec:opti}}

\begin{figure}[b]
	\centering
	\includegraphics[width = 0.7\columnwidth]{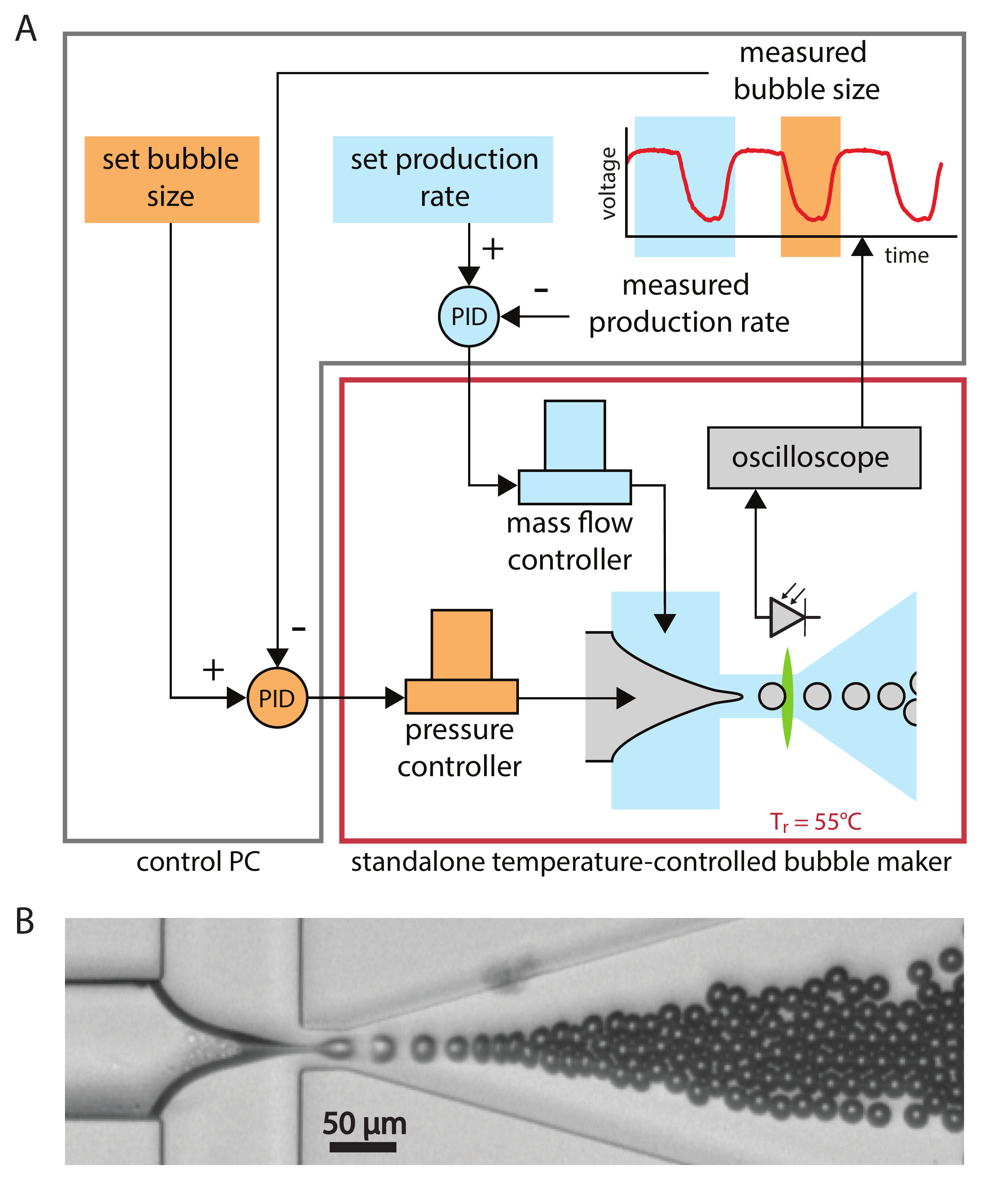}
	\caption{(A) Schematic representation of the active feedback loop for a liquid flow rate and gas pressure driven system. Two PID controllers drive the liquid flow rate and gas pressure until the set bubble size and production rate is reached. (B) Bright-field microscopy recording of the flow-focusing device producing phospholipid-coated microbubbles.}
	\label{fig:loop}
\end{figure}

Variations in bubble size can be caused by external parameters such as vibrations, dislodging dirt (a common issue in microfluidics) or aging of the chip which potentially results in a change in wetting properties of the channel walls resulting from, e.g., contact angle changes of plasma-treated PDMS~\cite{Kim2004c} or surfactant adsorption to the channel walls.~\cite{Kwiecinski2019} These disturbances decrease the production stability and thereby degrade the bubble size distribution both on short time scales as well as on the long-term.
In the experiments presented before (see Fig.~\ref{fig:IO}), the size of the produced bubbles changes for hundreds of seconds after a step in flow rate or gas flow rate. An important reason for this slow response is the lag in the gas pressure caused by dead volume and a continuous gas flow rate. Bubble production size and production rate will keep changing until the pressure has reached its new equilibrium, which may take several minutes. When implementing a feed-back control, the precision of the controller becomes second to the response time. To improve the response time, the gas mass flow controller has been replaced by a pressure controller.

\begin{figure}[htb]
	\centering
	\includegraphics[width = 0.8\columnwidth]{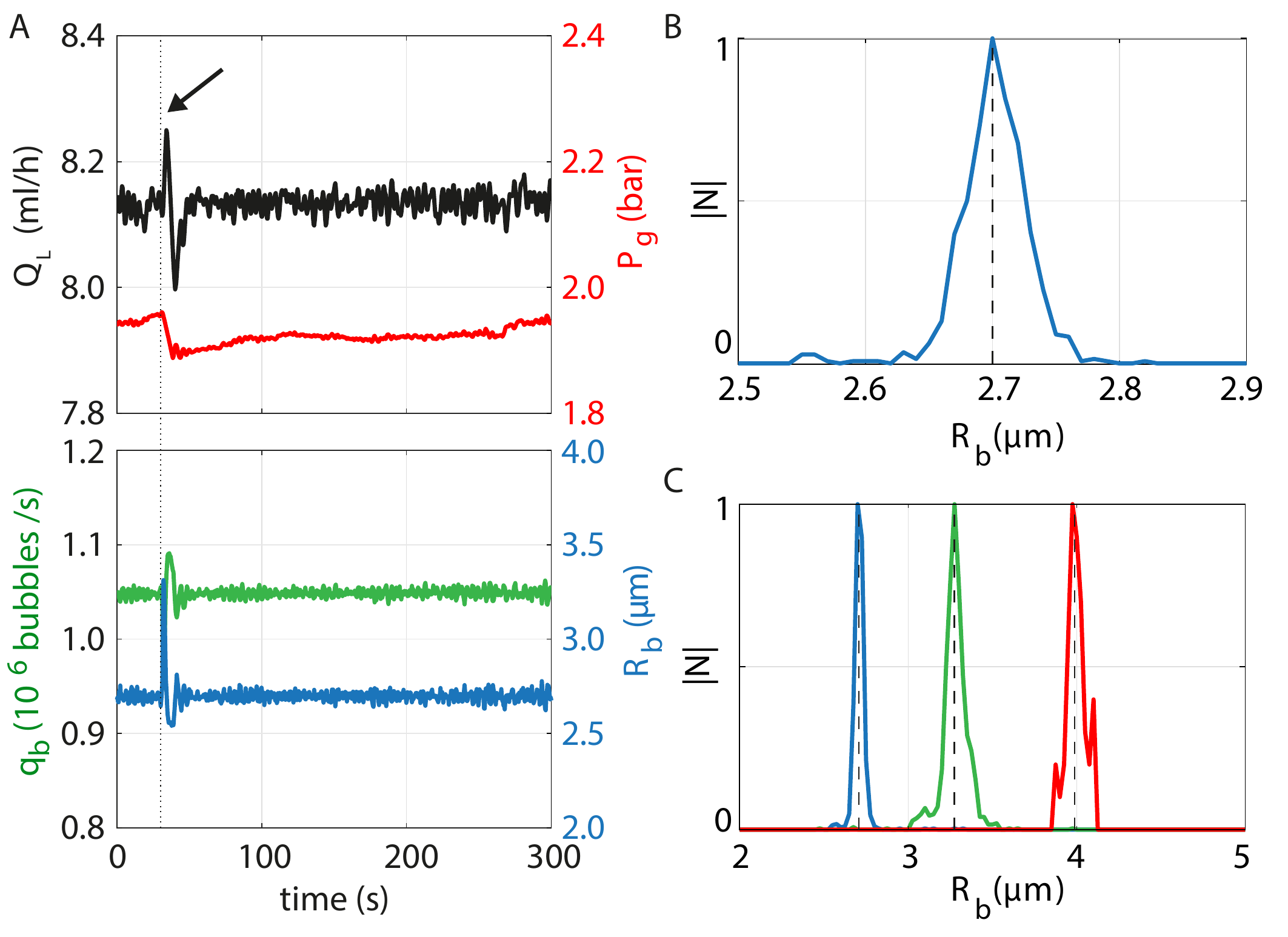}
	\caption{PID controlled stable bubble production. (A) The input (top) and output (bottom) of the microfluidic chip at a set production rate of 1.05~million bubbles/s and a bubble radius of 2.7~$\mu$m. The inputs are the liquid flow rate and the gas pressure. The outputs are the production rate and bubble size. After 30 s a disturbance occurs (see arrow), the input changes and the output stabilizes to its set rate and size within 10 s, even though the input changes for a duration of 200~s. (B) On-chip size distribution of the bubbles measured in Fig.~\ref{fig:stable}A. The dotted line is the set on-chip radius of 2.7~$\mu$m. (C) On-chip size distributions of different bubble populations produced at set radii of 2.7, 3.2, and 4.0~$\mu$m (dotted lines).}
	\label{fig:stable}
\end{figure}

As illustrated in Fig.~\ref{fig:loop}A, the direct read-out of the bubble size and of the production rate can be used in a feedback loop and stabilized by a classic PID controller to control and stabilize precisely and quickly the output of the flow-focusing device. A bright-field example of the flow focusing device producing phospholipid-coated bubbles at high speed is shown in Fig.~\ref{fig:loop}B. This operation can be performed independently of the microfluidic chip geometry used. This strategy uses the full capability of the system and reduces the transient times to a minimum by removing the explicit constraints on the flow and pressure controllers by placing them on the output of the chip. The stability is now only limited by the read-out and processing speed of the waveforms. Overall, an active feedback renders the system easier and more intuitive to operate.

\begin{figure}[b]
	\centering
	\includegraphics[width = 0.8\columnwidth]{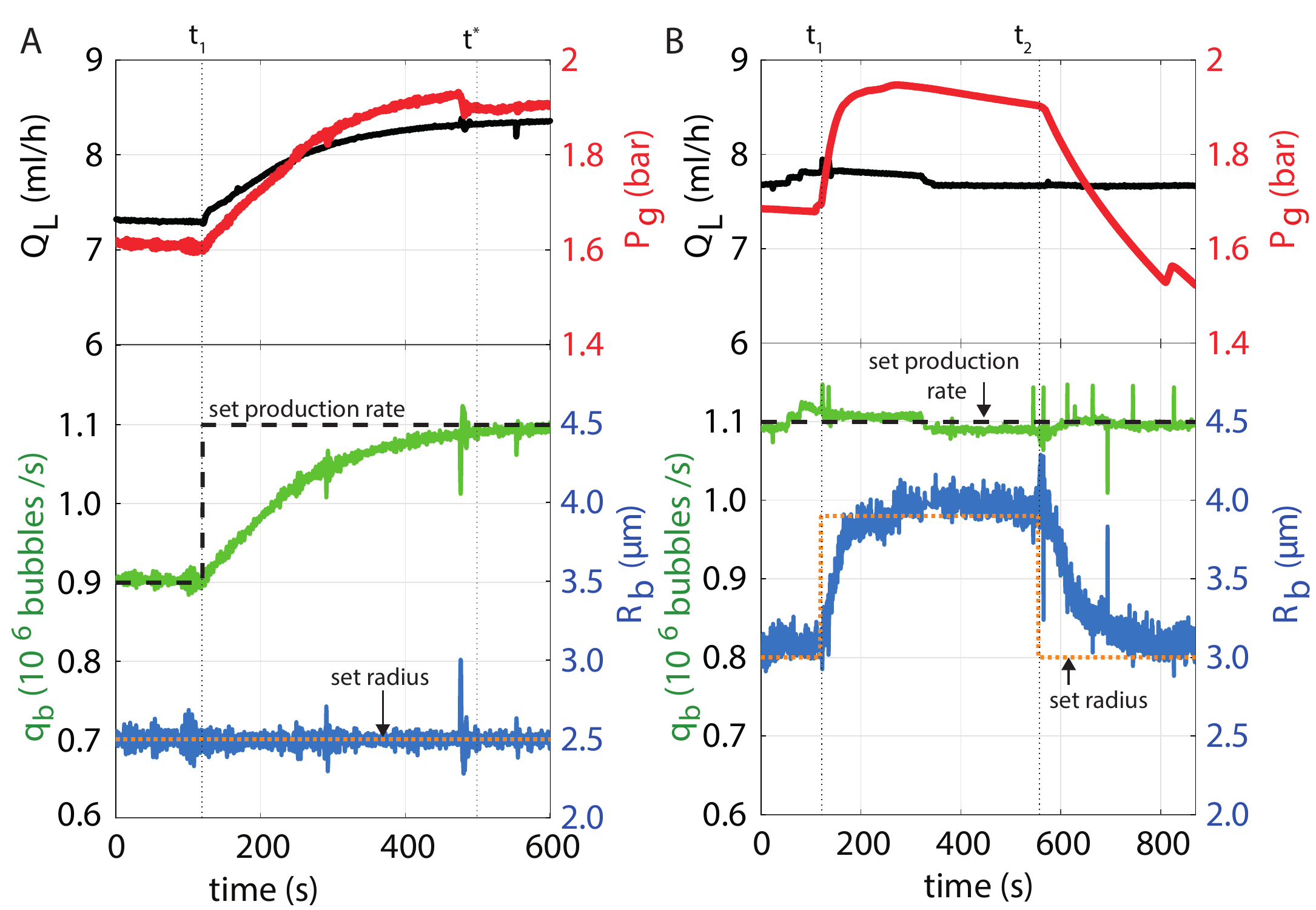}
	\caption{User-selected production rate and bubble size. (A) Response of the system to a change in set production rate from 0.9$\times$10$^6$ to 1.0$\times$10$^6$~bubbles/s at $t_1$. Both the liquid flow rate and the gas pressure increase to adjust the production rate while maintaining a constant bubble radius of 2.5~$\mu$m. At $t_2$ the set production rate is reached and the inputs stabilize. (B) The response of the system to a change of the set radius from 3.0~$\mu$m to 3.9~$\mu$m and back to 3.0~$\mu$m at $t_1$ and $t_2$, respectively. First the gas pressure is increased until the larger bubble size is reached. When at~$t_2$ the set bubble size is lowered, the gas pressure drops. The set production rate was kept constant at 1.1$\times$10$^6$~bubbles/s.}
	\label{fig:feedback}
\end{figure}

Figure~\ref{fig:stable}A demonstrates the stability of the feedback loop proposed in Fig.~\ref{fig:loop}A and implemented in our system. After 30~s, a sudden change in the flow properties of the chip (e.g. caused by a dust particle) disturbs the ongoing bubble production in terms of both, production rate (green curve, bottom panel) and bubble size (blue curve, bottom panel). The system immediately responds by changing the liquid flow rate (black curve, top panel) and the gas pressure (red curve, top panel). Within 10~s, the bubble production is back to the requested parameters. The constant drop in gas pressure after this event further indicates that this event is not a temporary disturbance but requires a long-lasting and active correction (over 3~min).
The size distribution calculated from Fig.~\ref{fig:stable}A is plotted in Fig.~\ref{fig:stable}B. Despite the disturbance, the fast correction ensures a narrow size distribution over time around the set radius of 2.7~$\mu$m. Size distributions of 3~bubble suspension produced at different set radii of 2.7, 3.2, and 4~$\mu$m are shown in Fig.~\ref{fig:stable}C.

Besides active stabilization, the feedback system offers the possibility to synthesize bubbles at user-specified rates and sizes as shown in Fig.~\ref{fig:feedback}. Figure~\ref{fig:feedback}A shows the response of the system after changing the requested production rate from 0.9$\times$10$^6$~bubbles/s to 1.1$\times$10$^6$~bubbles/s at $t_1=120$~s. Both the liquid flow rate and gas pressure go up to increase the production rate while maintaining a constant bubble size. Figure~\ref{fig:feedback}B shows the complementary response of the system upon changing the requested bubble radius from $3.0~\mu$m to $3.9~\mu$m at $t_1 = 120$~s. When the larger bubbles are requested, the gas pressure increases for approximately 1~minute until the production stabilizes at the new bubble size. The liquid flow rate slightly compensates to maintain the production rate. The small correction required here is in line with Fig.~\ref{fig:freq_and_bub_si} where we show that the gas flow rate indeed has a minor effect on the production rate. Note that in Fig.~\ref{fig:feedback}A the system also actively compensates for a sudden change at $t_2 = 480$~s while still adapting the chip output to comply with the user-requested parameters. 

Numerous chip designs are available in literature, and many more flow regimes are used to produce microbubbles or droplets. Their relative efficiency is very often characterized in terms of throughput and a polydispersity index (PDI). The PDI is typically measured in the final product, however it can also be directly measured within the chip using a high-speed camera. Such a camera allows for the determination of a PDI a posteriori using (intense) image processing techniques. The analysis is, however, restricted to a few hundred bubbles produced in a row. The PDI is taken as the standard deviation (or $1/e$ half width) of the size distribution assuming a Gaussian distribution. The outcome of such an analysis is often a PDI of less than $1\%$.~\cite{Hettiarachchi2007,Shin:2012zr}
Fig.~\ref{fig:stable}A (bottom), however, shows that there is a variation in produced bubble size on the timescale of seconds. This is thus also a more relevant timescale to characterize a PDI. To illustrate the discrepancy, the PDI evaluated on short times (milliseconds) from Fig.~\ref{fig:stable}A is lower than $1\%$. However, it jumps to up to $3\%$ when considered on a relevant duration (seconds), see the histogram of Fig.~\ref{fig:stable}B. This point is crucial not only for the characterization of a microfluidic technology or operational regime, but also to form a realistic characterization of the suspension produced by the microfluidic chip.~\cite{Talu2008} For example, polydispersity is one of the main measures to quantify the aging of a bubble/droplet population. A real-time system that can monitor continuously is thus a more reliable way to obtain an accurate estimate of the PDI.

Rather than going into specific details of the response of the particular chip used in the present work, we aimed at providing equations that facilitate a general description, applicable to a range of different microfluidic chip geometries. Therefore, in the present analysis, the considerations were kept simple and general, without sacrificing the essence of the physics. The weak response in production rate to a varying gas flow is surprising and interesting for further study. While the laws extracted for the production rate are not universal, the strategy can be easily adapted to different chip designs and different flow regimes. For the sake of simplicity, a choice was also made to use the ideal gas law. This assumption may give some error for high-molecular weight gases such as perflurorocarbons (typically used for ultrasound contrast agent microbubble production). However, we do not expect this effect to be significant as the proposed method uses the change in volume and pressure rather than the absolute relation.

In Sec.~\ref{sec:opti}, we have shown the response of the system when using a pressure controller. It must be noted that the use of a pressure controller creates, by default, a nonlinear response depending on whether the pressure is increased or decreased. In the former case, the response can be as fast as the controller itself if the system does not contain an excessive dead volume but in the latter case, the response is much slower since the excess pressure has to dissipate through the chip. This issue can be remedied by installing a small leak valve between the pressure controller and the chip that will allow for the pressure to drop more quickly. A leak valve is however not ideal when using high-molecular weight (greenhouse) gases such as perfluorocarbons that are typically expensive. 

\begin{figure}[b]
	\centering
	\includegraphics[width = .55\columnwidth]{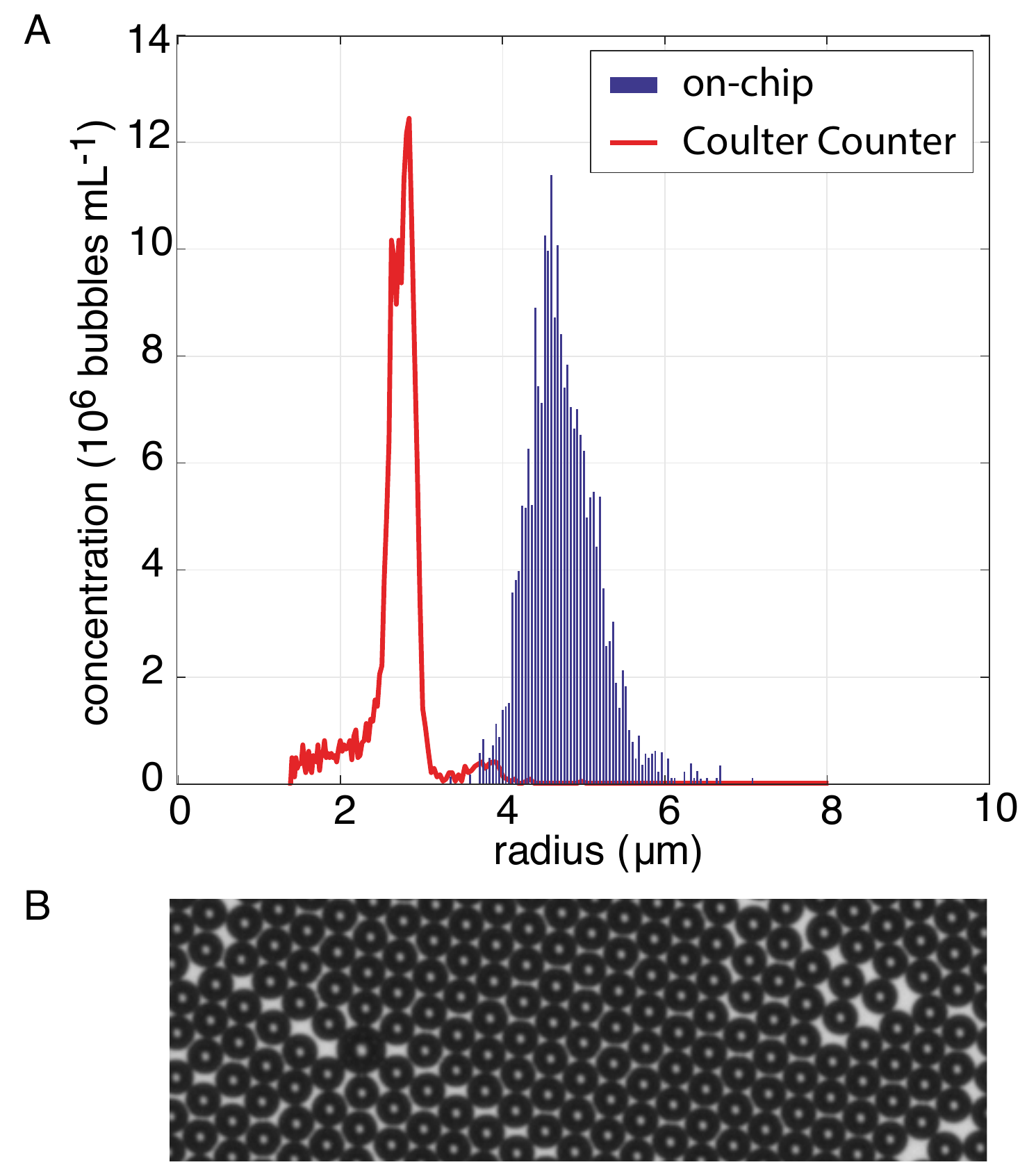}
	\caption{(A) Comparison between an on-chip bubble size distribution measured with the optical system (blue histogram) and the corresponding size distribution of the same bubble suspension after the bubbles have reached their stable size (red curve) measured using a Coulter Counter. The size decrease results from bubble dissolution as a result of the loose packing of the lipid monolayer around the freshly formed bubbles (B) Optical microscope image of the stable phospholipid-coated bubbles. The hexagonal packing confirms that the bubbles are highly monodisperse.}
	\label{F:10}
\end{figure}

%%%%%%%%%%%%%%%%%%%%%%%%%%%%%%%%%%%%%%%%%%%%%%%%%%%%%
Microfluidically-produced lipid-coated bubbles initially form with a loosely packed lipid shell.~\cite{Segers2016} The corresponding high surface tension drives bubble dissolution which mechanically compresses the lipid monolayer shell until the final stable bubble size is reached with near-zero surface tension. The on-chip to final size ratio depends on the lipid mixture and the addition of co-solvents to the aqueous phase.~\cite{Segers2016,Segers2017,Segers2019} For the present phospholipid mixture (DSPC/DPPE-PEG5k, 9:1 molar ratio) we showed before that the bubble size decreases by a factor of 2.2 during stabilization.~\cite{Segers2019} Fig.~\ref{F:10}A shows the on-chip size distribution as measured with the present optical detection system (blue histogram) next to the size distribution of the stable bubbles (collected as in~\citet{Segers2020}) that was measured using a Coulter Counter 15 min after bubble collection (solid red line). Figure~\ref{F:10}B shows a microscope image of the stable bubbles. The size reduction in Fig.~\ref{F:10}A amounts approximately to the expected dissolution ratio. Furthermore, the ratio between bubble concentrations off- and on-chip (production yield) may decrease after bubble pinch-off through on-chip coalescence and foam formation. However, when the present lipid mixture is used together with the appropriate gas mixture, the bubble concentrations on- and off-chip are identical.~\cite{Segers2020} Therefore, using the available knowledge on bubble stabilization,~\cite{Segers2017,Segers2019} foam formation,~\cite{Segers2020} and coalescence,~\cite{Segers2017,Segers2019} the present measurement method does provide concentration and size data representative of the stable bubble suspension.

Here, we have focused on the production of bubbles because of our intended application. However, the same technology can be used for the production of droplets. In general, producing droplets using microfluidics is significantly easier than producing bubbles owing to incompressibility of the medium and increased long-term stability of the droplets. Micron-sized droplets are also mostly unaffected by dissolution. The only relevant change to the system when using droplets is therefore the need for an intermediate pressurized container that stores the liquid used for the inner phase.

We have made the choice to regulate independently the production rate and the bubble size to demonstrate the potential of the proposed method. In practice, control over the bubble or droplet size is much more important than controlling the production rate. A more robust version of the system can be obtained by setting a constant gas pressure and by making a feedback loop between the measured bubble size and the liquid flow rate. This approach was also tested (data not shown), requires no major change to the system, and improves the response time and precision while releasing control over the production rate.

\section{Conclusions}
A bubble counter and sizer is proposed, based on the optical transmission of laser light through a microfluidic chip. We have successfully measured production rates of 10$^6$ bubbles/s with bubble sizes of the order of a few micrometers. The method only uses basic knowledge of microfluidics, which makes it applicable to a multitude of chip designs and platforms. It does not require any intrusive features to be implemented in the design of the chip itself, the requirement being optical transparency, a condition that is typically met in current lab-on-a-chip designs.
We also show that this real-time measurement can be used in a feedback configuration to control, stabilize and optimize the functionality of the microfluidic platform. It has been successfully implemented in our stand-alone device, controlled simply by a laptop. The proposed method can promote the use of microfluidic technologies to the inexperienced user and therefore opens a route to industrial and bedside production of clinical-grade ultrasound contrast agents.

\section*{Acknowledgements}
The authors gratefully acknowledge financial support from Bracco Suisse S.A. This work is supported by NanoNextNL, a micro and nanotechnology consortium of the Government of the Netherlands and 130 partners. The authors are grateful to Elena de Castro Hernandez for stimulating discussions. We also warmly thank Martin Bos and Bas Benschop for their technical support and Frans Segerink for his help with the optical sensing and associated electronics.

\section*{References}
\bibliography{BMpaper_arxiv}

\section{Appendices}
\subsection{Gas pressure and transient flows}
To characterize the bubble production we start by considering the properties of the on-chip fluid flows. Typically, microfluidic flows are laminar as can be shown from the low Reynolds number $R_e = {\rho_L \bar{u}_L \ell}/{\mu_L}$, where $\bar{u}_L$ is the average flow velocity, $\ell$ the typical length scale of the flow, here chosen to be the hydraulic diameter (4 times the area over the perimeter) of the smallest channel, $\rho_L$ is the density of the liquid and $\mu_L$ the viscosity of the liquid. During the experiments the average flow velocity never exceeded 10~m/s, the hydraulic diameter was $\sim$18~$\mu$m and water was used resulting in a maximum Reynolds number of 180, corresponding to a laminar flow.~\cite{Rands:2006kx,Mohiuddin-Mala:1999uq,Morini:2004fk}
\\
Bubble formation is driven by the gas pressure at the focusing region of the chip where the gas and liquid meet. The gas pressure in the supply line can be considered homogeneous because of the low viscosity of the gas, which results in negligible viscous dissipation. The ideal gas law and mass conservation can therefore be used to calculate the rate of change of the gas pressure~$P_g$, as follows:
\begin{equation} \label{eq:masscons}
\dot{P_g} = \frac{1}{V_g} \left( \frac{T_r R_g}{\tilde{M_g}} \dot{m}_{g} - Q_{g} P_{atm} \right),
\end{equation}
with $V_g$ the volume of the gas line, $T_r$ the ambient temperature,  $\tilde{M_g}$ the molar mass of the gas, $R_g$ the ideal gas constant, $\dot{m}_{g}$ the gas mass flow rate at the inlet, $Q_{g}$ the gas volume flow rate at the outlet of the chip, and $P_{atm}$ the atmospheric pressure, see Fig.~\ref{fig:Parameters}A.
The pressure drop over the outlet channel of the chip is neglected since the outlet typically has a much larger cross-section than the rest of the chip geometry.
\\
The pressure drop over the narrow channel in which the bubbles are formed can be written as:~\cite{Judy:2002vn,Steinke:2006ys}
\begin{equation} \label{eq:genelossesinmicrochan}
\Delta P = P_g - P_{atm} = \tfrac{1}{2}{K_i\rho_L \bar{u}_L^2} + \tfrac{1}{2}{K\left(x\right)\rho_L  \bar{u}_L^2} + \tfrac{1}{2}{K_o\rho_L  \bar{u}_L^2}   + R_{H0}  \bar{u}_L S,
\end{equation}
where $K_i$ characterizes the entrance losses due to the reduction of the cross-section of the microfluidic channel, $K_o$ quantifies the losses due to the increase in cross-section of the outlet channel. $K\left(x\right)$ is the Hagenbach factor and characterizes additional losses due to the development of a flat velocity profile into a  parabolic Poiseuille profile.~\cite{Judy:2002vn,Steinke:2006ys} Finally, $S$ represents the cross-section of the bubble formation channel (250~$\mu$m$^2$ for the present chip) and $R_{H0}$ the hydraulic resistance, see Fig.~\ref{fig:Parameters}B. Across the channel in which the bubbles are formed (red box Fig.~\ref{fig:Parameters}A), the pressure drops, resulting in an increasing volume gas flow rate due to the compressibility of the gas. $\bar{Q}_{g}$ can then be written as an average gas flow rate which then needs to be quantified. The lower end (and more relevant) of the range of Stokes numbers:
\begin{equation} \label{eq:stokes}
St = \frac{\bar{u}_L {R_b}^2 \rho_L}{4.5	 \mu_L L} \approx 0.1 - 10 ,
\end{equation}
is sufficiently low to say that the bubble velocity will be identical to that of the surrounding liquid. In Eq.~\ref{eq:stokes}, $R_b$ is the bubble radius and $L$ is the length of the bubble formation channel. The average velocity then takes the shape:
\begin{equation} \label{eq:velocity}
\bar{u}_L = \frac{Q_{tot}}{S} = \frac{ Q_L + \bar{Q}_{g}}{S}.
\end{equation}Furthermore, the inertial pressure drop depends on the density of the fluid, here a mixture of gas and water. We therefore define the density:
\begin{equation} \label{eq:density}
\rho_m = \rho_L \frac{Q_L}{Q_L +\bar{Q}_{g}},
\end{equation}
with $\rho_L$ = $10^3$ kg/m$^3$ the density of the liquid phase. 
The precise expression of each of the factors in Eq.~\ref{eq:genelossesinmicrochan} depends on the geometry of the chip. Using Eq.~\ref{eq:velocity} and Eq.~\ref{eq:density}, we rewrite Eq.~\ref{eq:genelossesinmicrochan} in a more generic form:
\begin{equation} \label{eq:hydrores2}
\Delta P = P_g - P_{atm} = R_{H0} \left(  Q_L + \bar{Q}_{g}\right) + K_2 Q_L \left(Q_L + \bar{Q}_{g}\right),
\end{equation}
where $R_{H0}$ and $K_2 $ are the linear hydraulic resistance and the quadratic inertial loss coefficient of the chip, respectively. $R_{H0}$ and $K_2 $ can be calculated using the pressure measured in the gas line in relation with the set flow rates. 
\\
As noted before, an assumption is needed to express the average gas flow rate $\bar{Q_{g}}$. The change in the gas volume flow rate while crossing the channel requires a deeper knowledge of the local flow behavior, which is difficult to translate from chip to chip (and for the general application of this work undesirable). It is therefore convenient to write a simple approximate relation: 
\begin{equation} \label{eq:phiav}
\bar{Q}_{g} \approx \frac{Q_{g}}{2} \left( 1+\frac{P_{atm}}{P_g}  \right).
\end{equation}
We will show later that this expression is in fact sufficiently accurate for our purpose.
Combining Eq.~\ref{eq:masscons}, Eq.~\ref{eq:hydrores2} and Eq.~\ref{eq:phiav} in the limit of high pressure driving of the chip ($\bar{Q}_{g} \approx \frac{Q_{g}}{2}$) results in:
\begin{equation} \label{eq:Prestran}
\begin{split}
\dot{P_g} \approx  -\frac{2 P_{atm}}{V_g \left(R_{H0} + K_2 Q_L\right)}P_g +   \frac{T_r R_g}{\tilde{M_g} V_g} \dot{m}_g \\ + \frac{2 {P_{atm}}^2}{V_g \left(R_{H0} + K_2 
	Q_L\right)} + 2\frac{Q_L P_{atm}}{V_g},
\end{split}
\end{equation}
Eq. \ref{eq:Prestran} is a first-order differential equation that presents a typical transient time: 
\begin{equation} \label{eq:tauP}
\tau = \frac{1}{2} \frac{V_g \left(R_{H0} + K_2 Q_L\right)}{P_{atm}}.
\end{equation}
Note that the prefactor $1/2$ becomes 1 in case of low flow rates ($\bar{Q}_{g} \approx Q_{g}$), implying that a strongly driven chip will respond faster than a mildly driven chip.
From Eqs. \ref{eq:hydrores2} and \ref{eq:tauP}, one can determine an approximate solution for the response to a pressure change:
\begin{equation} \label{eq:app_pressure}
P_{g}\left(t\right) \approx P_{g,t=0} + \left(P_{g,eq} -P_{g,t=0} \right)\left( 1-e^{-t/\tau} \right),
\end{equation}
with
\begin{equation}\label{eq:app_pressure_eq}
P_{g,eq}  = \frac{1}{2} \left( \kappa + \sqrt{\kappa^2 + 2 \left(R_{H0} + K_2  Q_L\right) Q_L P_{atm}}\right),
\end{equation}
and where
\begin{equation}\label{eq:delt}
\kappa = P_{atm}+\left(R_{H0} + K_2 Q_L\right)\left( Q_L + \frac{T_r R_g \dot{m}_g}{P_{atm} \tilde{M_g}}\right).
\end{equation}
For a more precise characterization, i.e., an in-detail study of a new chip design, the proposed reasoning can be repeated with added details on the flow dynamics.
One can then easily translate Eq.~\ref{eq:masscons} and Eq.~\ref{eq:hydrores2} to the gas flow rate:
\begin{equation} \label{eq:phiouttran}
\dot{Q_{g}} \approx  -\frac{2 P_{atm}}{V_g \left(R_{H0} + K_2 Q_L\right)}Q_{g} +   \frac{2 T_r R_g}{\tilde{M_g} V_g \left(R_{H0} + K_2 Q_L\right)} \dot{m}_g - 2 \dot{Q_L}.
\end{equation}
The gas flow rate thus experiences the same transient as the gas pressure. It also becomes apparent that the gas flow rate depends in a non-linear way on the variations of the liquid flow rate. Thus, a constant input gas flow rate does not necessarily imply a constant outlet gas flow rate.
\\
\subsection{Bubble sizing}
In contrast to the production rate, the bubble radius $R_b$ is not directly measured. The bubble radius is a function of both the passing time of the bubble $T_{pass}$ (Fig.~\ref{fig:setup}B) measured from the waveforms, and the bubble velocity. The passing time of the bubbles is measured as follows. First, the DC component of the waveform containing several bubble shadows was subtracted from the waveform. Every passing bubble corresponds a decrease in intensity, therefore each bubble relates to two zero crossings in the waveform. At these zero crossings the local derivative was then taken. The intersections of these two slopes with the maximum voltage correspond to the entrance and exit time of the bubble into the laser sheet. The maximum voltage was determined by finding the threshold that contained 95\% of the datapoints to be robust against noise. {In the laminar regime, flow develops from flat to parabolic over a typical entrance length $\ell_e$ given by:~\cite{Everts2020}}
\begin{equation} \label{eq:entrance_length}
\ell_e \approx 6.10^{-2} R_e D_h = 6.10^{-2} \frac{Q {D_h}^2 \rho_L}{\mu_L S}.
\end{equation}
In the chip presently investigated ($\ell_e \approx 200~\mu$m), the flow is thus not fully developed at the end of the bubble formation channel ($L=30~\mu$m), where the laser sheet was positioned. If we assume that the bubbles do not to significantly influence the overall flow profile and that they are positioned in the center of the flow profile, the bubble velocity is:
\begin{equation} \label{eq:entrance_length_2}
V_b = 2 \beta \bar{V},
\end{equation}	
where $\beta$ is a flow profile parameter added to account for the undeveloped flow and it can vary between $\beta = 1$ for a parabolic profile and $\beta = 1/2$ for a flat profile. Therefore:
\begin{equation} \label{eq:bub_si_2}
R_b = - \dfrac{D_{laser}}{2} + \beta T_{pass} \frac{Q_{g}+Q_L}{S},
\end{equation}
with $D_{laser}$ the thickness of the laser sheet, $T_{pass}$ extracted from the waveforms, and $Q_L$ from the liquid mass flow rate measured directly by the mass-flow controller. $Q_g$ follows from the set gas mass flow rate taking into account the gas pressure. Eq.~\ref{eq:hydrores2} can be rewritten in the following form:
\begin{equation} \label{eq:phioutandB}
Q_{g} = \frac{2}{1+\frac{P_{atm}}{P_g} }\left( \frac{P_g - P_{atm}}{R_{H0} + K_2 Q_L} - Q_L \right).
\end{equation}
where $P_g$ follows directly from Eq.~\ref{eq:app_pressure_eq} established before.
Finally, $D_{laser}$ can be estimated from the diffraction limit:\cite{Born1999}
\begin{equation} \label{eq:las_diam}
\dfrac{D_{laser}}{2} \approx 1.22\frac{\lambda F}{D} \approx 0.8~\mu \text{m}.
\end{equation}
Here, $\lambda = 405$ nm is the wavelength of the laser, $F$ is the working distance of the objective and $D$ the diameter of the beam entering the microscope. This result can easily be translated to droplet sizing by considering the inner flow to be incompressible, which reduces the prefactor of Eq.~\ref{eq:phioutandB} to 1 since in that case $\bar{Q}_{g} = Q_{g}$.
\\
Note that a more simple approach consists in writing:
\begin{equation} \label{eq:qg_practical}	
Q_{g} =  \frac{4}{3} \pi {R_b}^3 f_b,
\end{equation}
\\
where $f_b$ is the microbubble production frequency, which is directly measured from the waveforms. Eq.~\ref{eq:bub_si_2} then becomes a polynomial of order 3 in $R_b$ that can be solved to recover the bubble radius. This approach, however, can give an ambiguous result as there is no unique solution owing to the existence of multiple roots.

\end{document}